\documentclass[aps,prd,nofootinbib,twocolumn,superscriptaddress,preprintnumbers,balancelastpage,longbibliography]{revtex4-2}
\usepackage{aas_macros}

\usepackage[colorlinks=true,linkcolor=blue,citecolor=blue,urlcolor=blue]{hyperref}
\usepackage{orcidlink}
\usepackage{placeins}
\usepackage{lipsum}

\usepackage{graphicx}
\usepackage[normalem]{ulem}
\usepackage{amssymb}

\usepackage{graphicx}
\usepackage{dcolumn}
\usepackage{bm}

\usepackage{lineno}
\def\lsim{\mathrel{\raise.3ex\hbox{$<$\kern-.75em\lower1ex\hbox{$\sim$}}}}
\def\gsim{\mathrel{\raise.3ex\hbox{$>$\kern-.75em\lower1ex\hbox{$\sim$}}}}

\begin{document}

\makeatletter
\typeout{Column width: \the\columnwidth}
\typeout{Text width: \the\textwidth}
\makeatother


\title{Dark Matter, Baryon Number, and Cosmic-Ray Antinuclei}

\author{Caleb Gemmell\orcidlink{0000-0002-6505-8559}}
    \email{cgemmell2@wisc.edu}
\affiliation{
    Department of Physics, Wisconsin IceCube Particle Astrophysics Center, University of Wisconsin, Madison, Wisconsin 53706, USA
}

\author{Dan Hooper\orcidlink{0000-0001-8837-4127}}
    \email{dwhooper@wisc.edu}
\affiliation{
    Department of Physics, Wisconsin IceCube Particle Astrophysics Center, University of Wisconsin, Madison, Wisconsin 53706, USA
}
\author{Seth Koren\orcidlink{0000-0003-2409-4171}}
    \email{skoren@nd.edu}
\affiliation{
    Department of Physics and Astronomy, University of Notre Dame, South Bend, Indiana 46556, USA
}
\author{Fabrizio Vassallo\orcidlink{0000-0002-0541-5606}}
    \email{fevassallo@wisc.edu}
\affiliation{
    Department of Physics, Wisconsin IceCube Particle Astrophysics Center, University of Wisconsin, Madison, Wisconsin 53706, USA
}

\begin{abstract}
Antideuterons and antihelium nuclei in the cosmic-ray spectrum have long been considered a smoking gun signature of dark matter annihilation, making the tentative observation of several such events by AMS highly intriguing. Conventional dark matter models, however, can produce only up to $\mathcal{O}(1)$ antideuteron events at AMS and are not capable of generating observable fluxes of antihelium. In this Letter, we propose a class of models in which dark matter annihilates into particles carrying baryon and lepton number, whose subsequent decays produce enhanced fluxes of antinucleons and antinuclei. Such scenarios are motivated by Grand Unified Theories and can lead to an order-of-magnitude or larger enhancement in the resulting antideuteron and antihelium-3 fluxes, providing a means by which to potentially explain the events reported by the AMS Collaboration. 
\end{abstract}

\maketitle


\textit{Introduction}---While astrophysical mechanisms typically generate more matter than antimatter in cosmic rays, the processes of dark matter annihilation and decay generally produce equal quantities of particles and antiparticles, motivating searches for antimatter in the cosmic ray spectrum~\cite{Silk:1984zy,Ellis:1988qp,Stecker:1985jc}. Indirect searches for dark matter involving cosmic-ray positrons~\cite{Turner:1989kg,Kamionkowski:1990ty,Baltz:1998xv,Hooper:2004bq,Cholis:2008hb,Arkani-Hamed:2008hhe}, antiprotons~\cite{Bergstrom:1999jc,Cuoco:2016eej,Giesen:2015ufa,Cholis:2019ejx}, and antinuclei~\cite{Chardonnet:1997dv, Donato:1999gy, Baer:2005tw, Brauninger:2009pe, Cui:2010ud, Carlson:2014ssa, Cirelli:2014qia} have each been extensively studied and pursued. 

The current generation of cosmic-ray experiments offers unique opportunities to search for signatures of dark matter annihilation~\cite{vonDoetinchem:2020vbj}. Intriguingly, the AMS-02 experiment~\cite{Battiston:2007aj} aboard the International Space Station has reported preliminary evidence for several cosmic-ray antideuterons in its dataset~\cite{vonDoetinchem:2020vbj}, along with \mbox{$\sim 10$} \mbox{antihelium-3} nuclei, and even a few antihelium-4 events~\cite{Ting:2023abc, Zuccon:2024abc}. Given that standard astrophysical processes are expected to produce only $\mathcal{O}(1)$ antideuteron and $\mathcal{O}(0.1)$ antihelium-3 events over 15 years of AMS exposure, new physics would appear to be required to accommodate this remarkable signal~\cite{Poulin:2018wzu,Duperray:2005si, Shukla:2020bql,DeLaTorreLuque:2024htu}. Even conventional models of dark matter annihilation would be unable to generate as many antihelium events as have been reported by AMS~\cite{Korsmeier:2017xzj, Poulin:2018wzu, Ding:2018wyi, DeLaTorreLuque:2024htu, DiMauro:2025vxp, Herms:2016vop, Cholis:2020twh, Kounine:2011bkq}. In addition, the balloon-borne General Antiparticle Spectrometer (GAPS) experiment completed its first flight in January 2026, providing a new and complementary probe of antideuterons~\cite{Aramaki:2015laa} and antihelium~\cite{GAPS:2020axg} in the cosmic-ray spectrum.

In light of this situation, a number of exotic scenarios involving dark matter~\cite{Coogan:2017pwt, Heeck:2019ego, Winkler:2020ltd, Winkler:2022zdu, Fedderke:2024hfy, Korwar:2024ofe, DiMauro:2026owr} or other new physics~\cite{Poulin:2018wzu, Bykov:2023nnr} have been proposed in an effort to explain the \mbox{antinuclei} events reported by AMS~\cite{Coogan:2017pwt, Heeck:2019ego, Cholis:2020twh, Winkler:2020ltd, Winkler:2022zdu, Fedderke:2024hfy, Korwar:2024ofe, DiMauro:2026owr}. In this  Letter, we consider a class of models in which the process of dark matter annihilation produces particles that carry baryon number, guaranteeing the production of antibaryons and enhancing the yield of antinuclei relative to conventional dark matter scenarios. Such models can provide a means of potentially accounting for the events reported by AMS while remaining consistent with existing constraints. 


\textit{The model}---There exist well-motivated extensions of the Standard Model which feature particles that carry net baryon number but no QCD color. Such particles can arise, for example, in models where baryon number is shared with the dark sector~\cite{Duerr:2013lka,Kaplan:2009ag,Shelton:2010ta,Elor:2018twp,Davoudiasl:2010am,Elahi:2021jia}, and in extensions where some combination of $U(1)_B$ and/or $U(1)_L$ are spontaneously broken global or gauge symmetries~\cite{Armbruster:2025jqs, FileviezPerez:2010gw,Kivel:2022rzr,Bittar:2024nrn,Barbieri:1981yr,Berezhiani:2015afa,FileviezPerez:2014lnj, Ma:2020quj}.
In the Standard Model itself, the gauge structure explicitly breaks each of $U(1)_B$ and $U(1)_L$ to $\mathbb{Z}_3$ subgroups \cite{Koren:2022bam,Wang:2022eag}, providing motivation for models in which these groups are also spontaneously broken by scalars with charge $3$~\cite{Armbruster:2025jqs,Koren:2022axd,Kivel:2022rzr,FileviezPerez:2010gw,Ma:2020quj}. 
In the context of Grand Unified Theories (GUTs), $U(1)_{B-L}$ is often gauged and embedded into a larger non-Abelian structure~\cite{Pati:1974yy,Fritzsch:1974nn,Allanach:2021bfe,Davighi:2022qgb,Delgado:2026a,Glashow:1984gc,Dong:2013wca, Gursey:1975ki,Davidson:1978pm,Mohapatra:1980qe}, resulting in new species with various values of baryon and lepton number. 

Concretely, we consider here a model in which dark matter particles, $\chi$, annihilate into a pair of scalars, $\phi \overline{\phi}$, which carry baryon and lepton number and antibaryon and antilepton number, respectively. We consider two representative charge assignments for the scalar: $\phi_{1,1}$ with $B=1,~L=1$, and $\phi_{3,3}$ with $B=3,~L=3$. In the $\phi_{1,1}$ case, the scalar decays to three quarks and one lepton. In contrast, the $\phi_{3,3}$ decays through three off-shell $\phi_{1,1}$ scalars to nine quarks and three leptons.

\begin{figure*}[t]
  \centering
  \begin{minipage}[c][5cm][c]{0.49\textwidth}
    \centering
    \includegraphics[height=2cm,keepaspectratio]{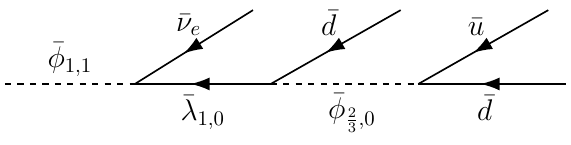}
  \end{minipage}\hfill
  \begin{minipage}[c][5cm][c]{0.49\textwidth}
    \centering
    \includegraphics[height=4.8cm,keepaspectratio]{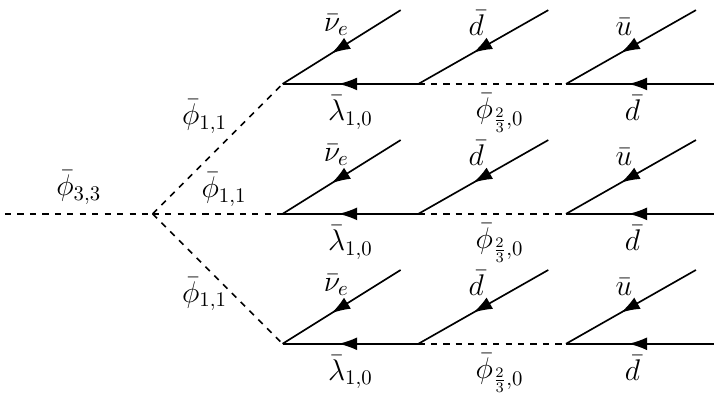}
  \end{minipage}
  \caption{In the models considered in this Letter, the dark matter annihilates to a pair of scalars with baryon and lepton number, $\chi \chi \rightarrow \phi_{i,j} \, \bar{\phi}_{i,j}$, where $i$ and $j$ denote the baryon and lepton number of the particle, respectively. These scalars then decay as shown here, for the cases of the models with $B=L=1$ (left) and $B=L=3$ (right).}
  \label{fig:feynman}
\end{figure*}

In Fig.~\ref{fig:feynman}, we show an example of how these scalar decays could proceed in a possible UV completion for the case of $\bar{\phi}_{1,1}$ (left) and $\bar{\phi}_{3,3}$ (right). Note that these processes each involve intermediate scalars, $\bar{\phi}_{\frac{2}{3},0}$, which carry QCD color and electric charge and are therefore required to have masses above the TeV scale to evade constraints from the Large Hadron Collider. Additionally, the decays also involve a colorless and electrically neutral fermion, $\lambda_{1,0}$, with $B=1,~L=0$, which could be much lighter. We take the intermediate particles to be heavier than the decaying scalar and integrate them out, modeling this process using higher-dimensional operators. For concreteness, we focus here on the following DM annihilation scenarios:  
\begin{equation}
    \chi\chi\rightarrow\phi_{1,1}\,\overline{\phi}_{1,1}\rightarrow (udd\nu_e)(\overline{u d d \nu}_e)
\end{equation}
and 
\begin{eqnarray}
    \chi\chi\rightarrow\phi_{3,3}\,\overline{\phi}_{3,3}\rightarrow 3(udd\nu_e)\,3(\overline{udd\nu}_e),
\end{eqnarray}
which correspond to operators of dimension 7 and 19, respectively. Even in this latter case, the scalar could potentially decay on a timescale smaller than that associated with propagation over Galactic length scales while respecting bounds from colliders.\footnote{This is possible despite LHC constraints in part due to the existence of a low-mass window for particles that decay to dijets which is poorly constrained in LHC analyses because large jet momentum cuts are needed to remove QCD backgrounds, as has been emphasized by Dobrescu \& Yu \cite{Dobrescu:2013cmh,Dobrescu:2021vak}. In our case the $\phi_{2/3,0}$ can be pair-produced so the leading bounds  \cite{ATLAS:2017jnp,CMS:2018mts} constrain $m_{\phi_{2/3,0}} > 520$ GeV. The uncolored intermediate particles $\lambda_{1,0}$ and $\phi_{1,1}$ are not produced copiously at the LHC, though there is a very weak bound on $m_{\lambda_{1,0}} \gtrsim $ few $\times$ GeV coming from $qq \rightarrow q \lambda$ production leading to a monojet plus missing energy signature, as recently analyzed in \cite{Hiller:2026osz}. If we consider the possibility of $N$ flavors of the uncolored $\lambda_{1,0}$ and $ \phi_{1,1}$ then for $N=1$, the $\phi_{3,3}$ decay length can be $\mathcal{O}(1 \text{ kpc})$ for $m_{\phi_{3,3}} \gtrsim 12$ GeV, while if we consider $N=3$ flavors mirroring the SM generation structure, such decay lengths are possible for all $m_{\phi_{3,3}} \gtrsim 5$ GeV.} As we show in Sec.~S3 of the Supplemental Material, our results do not significantly change if we were to consider decays to other combinations of quarks or leptons.


\textit{Event generation and antinuclei formation}---To calculate the multiplicity of antinuclei produced through the process of dark matter annihilation in the models considered here, we begin with the quarks and leptons produced in the decay of a scalar, $\overline{\phi}$. Since the decay occurs through off-shell processes, we sample the quark and lepton energies assuming a flat $n$-body phase space distribution in the massless-daughter limit, and distribute the momenta isotropically in the rest frame of the $\overline{\phi}$. While this treatment neglects the Lorentz structure of the off-shell fermions that have been integrated out, it provides a conservative estimate for the antinuclei, as it maximizes the separation between antiquark jets and thus final state antinucleons. Additionally, this leaves us relatively agnostic to possible alternative intermediate decay processes from $\overline{\phi}$ to the final state. We then use \textsc{Pythia 8.313}~\cite{Bierlich:2022pfr} to simulate parton showering and hadronization, employing the \textsc{Vincia} shower algorithm~\cite{Fischer:2016vfv} to include the effects of electroweak radiation. We adopt the \textsc{Pythia} parameters from Ref.~\cite{DiMauro:2025vxp}, selected to match data from LEP, ALICE, and ALEPH.\footnote{This choice of parameters also addresses a known issue in some standard configurations, namely that off-vertex decays of bottom hadrons can produce $^3\mathrm{\overline{He}}$ at artificially enhanced rates. We have independently verified the conclusion that production from $b$-hadron decay is subdominant to hadronization.}

We implement the formation of antinuclei in these events using a standard coalescence model \cite{Butler:1961pr, Kapusta:1980zz}. Specifically, an antinucleus is taken to form if every \mbox{antinucleon} pair in the candidate nucleus has a momentum difference $|\vec{p}_i - \vec{p}_j|$ $< p_{\mathrm{coal}}$ and spatial separation smaller than $|\vec{x}_i - \vec{x}_j| < d_{\mathrm{coal}}$, both measured in the pair's center-of-mass frame, where $p_{\mathrm{coal}}$ is a free parameter that is tuned to fit observational rates. We have taken the standard value of $d_{\mathrm{coal}}=3~\mathrm{fm}$, corresponding to the spatial extent of light nuclei, ensuring that antinucleons can form a bound state only if they are produced within a region comparable to the size of the nuclear wave function. For antideuterons, we adopt $p_{\rm coal}=219 \, {\rm MeV}$, which reproduces the $\mathrm{\overline{D}}$ multiplicity measured by ALEPH in $Z$-resonance hadronic decays~\cite{ALEPH:2006qoi}. Similarly, we take $p_{\mathrm{coal}} = 206 \ \mathrm{MeV}$ for $^3\mathrm{\overline{He}}$, based on the multiplicity measured by ALICE in $pp$ collisions at $\sqrt{s} = 13 \ \mathrm{TeV}$ \cite{ALICE:2021mfm} (for an overview and comparison of models, see Ref.~\cite{DiMauro:2024kml}). 

\begin{figure}[!t]
    \centering
    \includegraphics[width=\columnwidth]{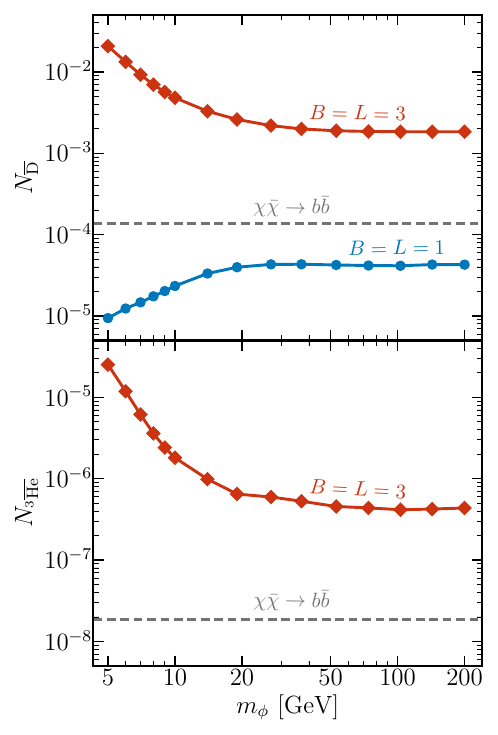}
    \caption{The average number of antideuterons (top) and antihelium-3 nuclei (bottom) per dark matter annihilation event in our models with $B=L=1$ or $B=L=3$ as a function of the mass of the decaying scalar, $m_{\phi}$. These results are compared to the yields predicted for $50$ GeV dark matter particles annihilating directly to $b\bar{b}$ (dashed line). The multiplicities of antideuterons and antihelium nuclei produced through dark matter annihilation can be enhanced by more than an order of magnitude in the $B=L=3$ model, especially for small values of $m_{\phi}$. Note that in the $B=L=1$ model, the number of predicted antihelium events is below the range shown in this figure.}
    \label{fig:antinuclei}
\end{figure}

\begin{figure}[!t]
    \centering
    \includegraphics[width=\columnwidth]{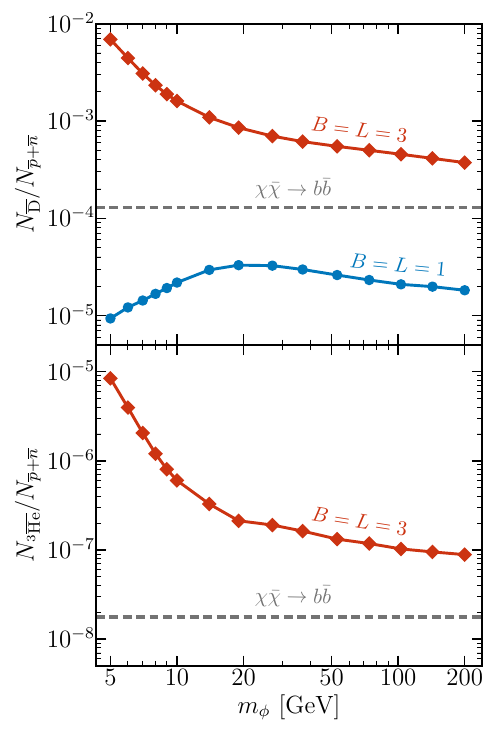}
    \caption{The ratio of antideuterons (top) and antihelium-3 nuclei (bottom) to antinucleons produced through dark matter annihilation in our models with $B=L=1$ or $B=L=3$ as a function of the mass of the decaying scalar, $m_{\phi}$. These results are compared to the yields predicted for $50$ GeV dark matter particles annihilating directly to $b\bar{b}$ (dashed line). The relative multiplicities of antideuterons and antihelium nuclei produced through dark matter annihilation can be enhanced by more than an order of magnitude in the $B=L=3$ model, especially for small values of $m_{\phi}$. Note that in the $B=L=1$ model, the number of predicted antihelium events is below the range shown in this figure.}
        \label{fig:antiprotons}
\end{figure}

\textit{Results}---In order for the process of dark matter annihilation to produce the $\sim 10$ antihelium-3 events reported by the AMS Collaboration, those annihilations must produce roughly $10$--$100$ times more \mbox{antihelium-3} nuclei than is predicted in conventional dark matter models~\cite{Poulin:2018wzu,Duperray:2005si, Shukla:2020bql,DeLaTorreLuque:2024htu} (although this factor could be reduced somewhat by adopting favorable cosmic-ray propagation parameters~\cite{Cholis:2020twh}). 

Our main results are presented in Fig.~\ref{fig:antinuclei}, where we show the multiplicity of antinuclei produced in the dark matter models considered in this study and compare it to a benchmark case of 50 GeV dark matter particles that annihilate directly to $b\bar{b}$. These results are sensitive to the value of the $\phi$ mass. In contrast, the mass of the dark matter itself does not impact the antinuclei multiplicity, although it will impact the spectrum of the resulting particles through the Lorentz boost of the scalar, $\phi$. This we plan to study in future work. 

For our model with $B=L=1$, the number of antideuterons and antihelium nuclei produced is actually slightly smaller than in the $\chi \chi \rightarrow b\bar{b}$ benchmark model. Although a larger number of antinucleons are formed in the $B=L=1$ model for most values of $m_\phi$ (see Sec.~S1 of the Supplemental Material), those particles are produced approximately isotropically in the rest frame of the decaying $\bar{\phi}$, suppressing the probability that they will coalesce into antinuclei. In contrast, the antinucleons that are produced in the conventional dark matter benchmark are more colinear, leading to somewhat greater antinuclei production. 

For our $B=L=3$ model, we find that the number of antideuterons and antihelium nuclei produced is enhanced by a factor of $\mathcal{O}(10$--$100)$ relative to the benchmark dark matter model. This enhancement results from two factors: (i) the larger number of antibaryons produced due to the scalar's baryon number, and (ii) the increased number of hadronic jets produced in the initial $\overline{\phi}$ decay. The larger number of antiquark jets both softens the momentum distribution of antinucleons and reduces their average spatial separation, increasing the likelihood of antinucleus formation. The larger antibaryon number of $\bar \phi_{3,3}$ ensures that each decay contains at least three antibaryons, greatly increasing the chance that an antinucleus can coalesce compared to the decays of $\bar \phi_{1,1}$ which may have as few as one antibaryon.

Because of the enhanced number of antinucleons that are formed through dark matter annihilation in the models presented here, constraints from the antiproton content of the cosmic-ray spectrum are expected to be more stringent than in conventional dark matter scenarios. In Fig.~\ref{fig:antiprotons}, we show the ratios of antideuterons and antihelium-3 nuclei produced per antinucleon in each of the models considered here. For the $B=L=3$ model, we find that this ratio can be significantly enhanced, especially for relatively small values of $m_{\phi}$. This allows these models to produce significantly enhanced fluxes of cosmic-ray antinuclei while remaining consistent with existing constraints from cosmic-ray antiprotons. 

\textit{Summary}---In this Letter, we have considered models in which dark matter annihilates into particles that carry baryon ($B$) and lepton ($L$) number. Such particles arise in a variety of well-motivated extensions of the Standard Model, including Grand Unified Theories. Because these particles carry baryon number, their decays generate an enhanced yield of nucleons and antinucleons. For models with $B=L=3$ or greater, dark matter annihilation can produce significantly larger multiplicities and fluxes of antideuterons and antihelium nuclei relative to conventional dark matter scenarios. This could allow dark matter annihilations to produce many of the cosmic-ray antideuteron and antihelium events that have been provisionally reported by the AMS Collaboration while remaining consistent with constraints from cosmic-ray antiprotons. Establishing whether these models can quantitatively reproduce the reported AMS-02 candidates will require a dedicated treatment of cosmic-ray propagation and detector response, as well as more detailed information on the antinuclei candidates than is currently publicly available. With GAPS having completed its first flight and AMS-02 continuing to collect data, our understanding of antinuclei in the cosmic-ray spectrum is likely to improve substantially in the coming years.

\bigskip

\textit{Acknowledgments}---We thank David Curtin for useful discussions. This work has been supported by the Office of the Vice Chancellor for Research at the University of Wisconsin-Madison, with funding from the Wisconsin Alumni Research Foundation. The work of S. K. is partially supported by the National Science Foundation under Grant No. PHY-2412701.


\bibliography{bibliography}

\clearpage

\appendix

\makeatletter
\typeout{Column width: \the\columnwidth}
\typeout{Text width: \the\textwidth}
\makeatother

\preprint{APS/123-QED}

\onecolumngrid
\begin{center}
  \textbf{\large Supplemental Material for Dark Matter, Baryon Number, and Cosmic-Ray Antinuclei}\\[.2cm]
  \vspace{0.05in}
  {Caleb Gemmell, Dan Hooper, Seth Koren, and Fabrizio Vassallo}
\end{center}

\setcounter{figure}{0}
\renewcommand{\thefigure}{S\arabic{figure}}

\setcounter{figure}{0}
\renewcommand{\thesection}{S\arabic{section}}

\onecolumngrid

\section{Average Antinucleon Yields per Dark Matter Annihilation}
\label{sec:S1}

\begin{figure}[h]
    \centering
    \includegraphics[width=4in]{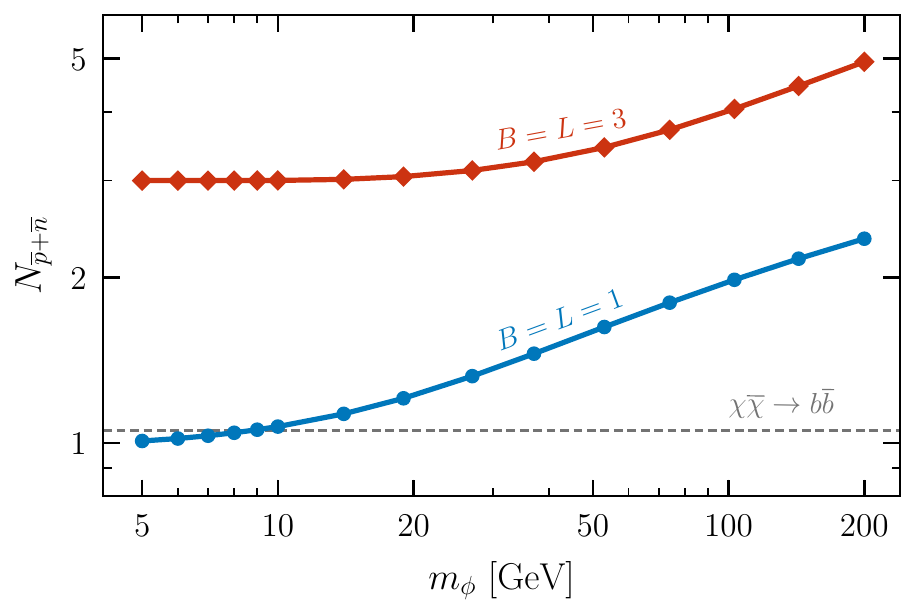}
    \caption{The average number of antinucleons predicted per dark matter annihilation event in our models with $B=L=1$ or $B=L=3$ as a function of the mass of the decaying scalar, $m_{\phi}$. These results are compared to the yields predicted for $50$ GeV dark matter particles annihilating directly to $b\bar{b}$ (dashed line).}
    \label{fig:antinucleons}
\end{figure}

In Fig.~\ref{fig:antinucleons}, we show the average number of antinucleons produced per dark matter annihilation event in our models with $B=L=1$ or $B=L=3$ as a function of the mass of the decaying scalar, $m_{\phi}$. These results are compared to the yields predicted for $50$ GeV dark matter particles annihilating directly to $b\bar{b}$. Note that the number of antinucleons produced always exceeds the baryon number of the decaying scalar. 

\section{Antinuclei Yields for Models with Various Baryon and Lepton Numbers}
\label{sec:S2}

\begin{figure}[h]
    \centering
    \includegraphics[width=0.49\columnwidth]{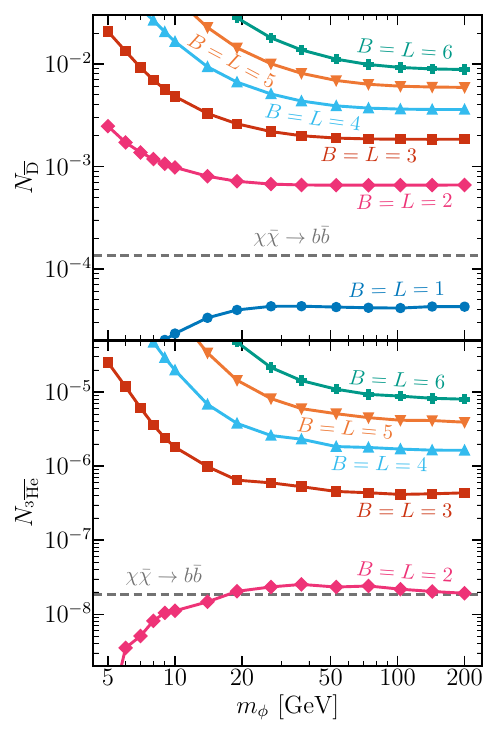}
    \includegraphics[width=0.49\columnwidth]{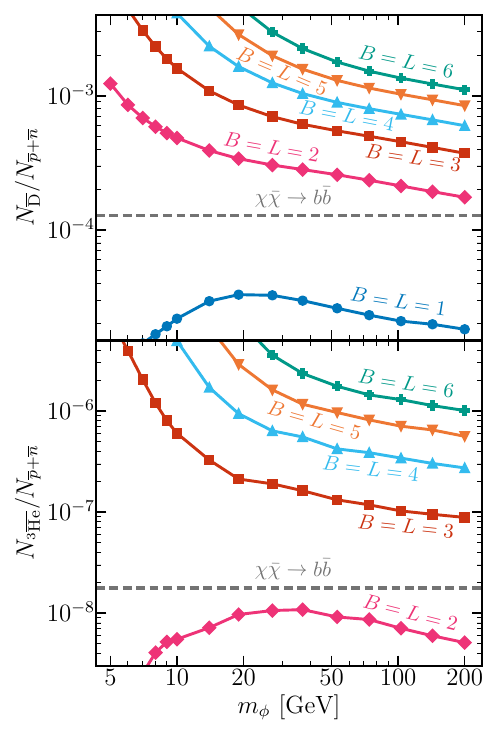}
    \caption{As in Figs.~2 and~3 in the main text, we plot the average antinuclei yield (left) and antinuclei-to-antinucleon ratios (right), for models in which the scalar $\phi$ carries different values of baryon and lepton number, up to $B = L = 6$.}
    \label{fig:uptosix}
\end{figure}

In Fig.~\ref{fig:uptosix}, we show the same results as in Figs.~2 and~3 of the main text, but for models in which the scalar, $\phi$, carries various values of baryon and lepton number, up to $B=L=6$. We find similar trends in $m_\phi$ across all $B,L$ values considered, with the enhancement growing with increasing $B,L$. We also find that when $B,L$ is less than the number of antinucleons in a desired antinuclei, its flux is suppressed relative to the benchmark dark matter model.

\section{Antinuclei Yields for Alternative Hadronic Decay Channels}
\label{sec:S3}

\begin{figure}[h]
    \centering
    \includegraphics[width=0.49\columnwidth]{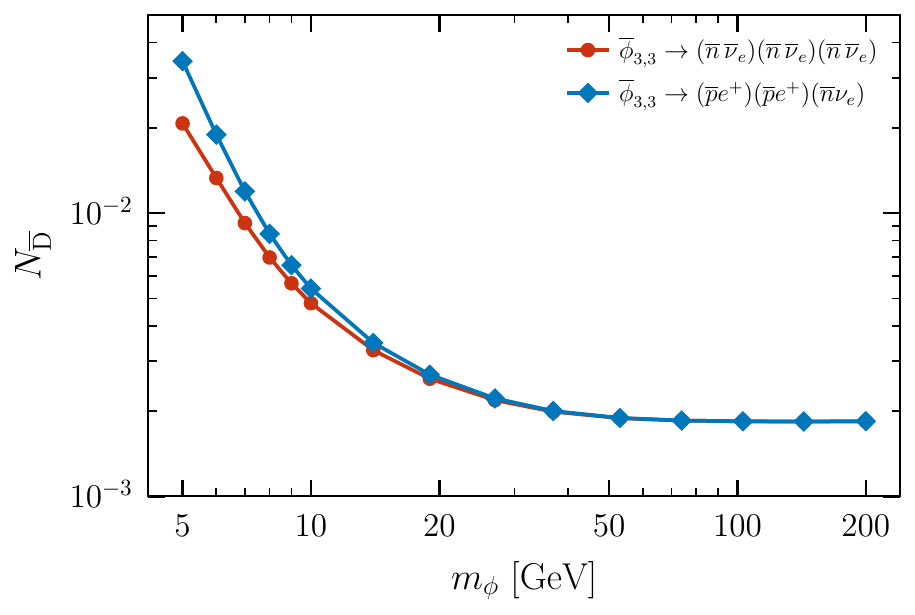}
    \includegraphics[width=0.49\columnwidth]{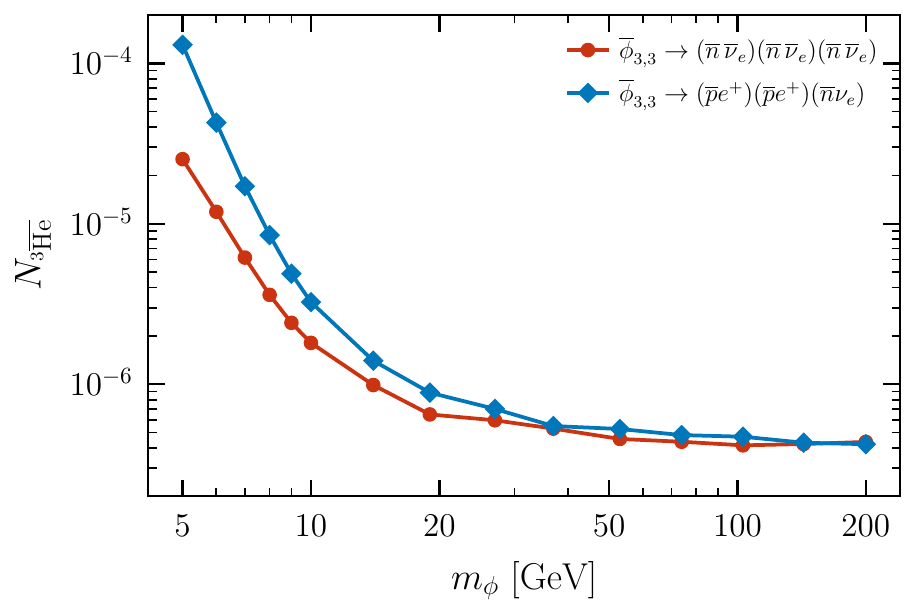}
    \caption{A comparison of the antideuteron and antihelium-3 yields for two different decay channels.}
    \label{fig:compare}
\end{figure}

In this {\it letter}, we have focused in the $B=L=3$ case and on decays of the form, $\overline{\phi}\rightarrow (\overline{n}\,\overline{\nu}_e)(\overline{n}\,\overline{\nu}_e)(\overline{n}\,\overline{\nu}_e)$. However, one might expect that fluxes of $^3\overline{\rm{He}}$ would be maximized for the decay channel, $\overline{\phi}\rightarrow (\overline{p}e^+)(\overline{p}e^+)(\overline{n}\,\overline{\nu}_e)$. In Fig.~\ref{fig:compare}, we compare the antideuteron and antihelium-3 yields between these two cases and find this is true only in the limit of small $m_\phi$. For larger masses, the two decay channels asymptote to the same rate. Thus, by considering the decay channel, $\overline{\phi}\rightarrow (\overline{n}\,\overline{\nu}_e)(\overline{n}\,\overline{\nu}_e)(\overline{n}\,\overline{\nu}_e)$, we have made conservative estimates for the resulting antinuclei yields. 

\end{document}